\documentclass{article}

\usepackage{graphicx} % Required for inserting images
\usepackage[utf8]{inputenc}
\usepackage[T1]{fontenc}
\usepackage{hyperref}
\usepackage{url}
\usepackage{times}
\usepackage{lmodern}
\usepackage{xcolor}
\usepackage{booktabs} 
\usepackage{amssymb}
\usepackage{amsmath}
\newcommand{\round}[1]{\ensuremath{\lfloor#1\rceil}}

\usepackage[style=apa,
citestyle=authoryear,
sorting=nyt,
sortcites=true,
autopunct=true,
babel=hyphen,
hyperref=true,
abbreviate=false,
backref=false,
dashed=false,
maxcitenames=2, 
apamaxprtauth=999, 
uniquename=false,
uniquelist=false,
backend=biber]{biblatex}

\title{Effects of survey design features on response rates: a meta-analytical approach using the example of crime surveys}

\author{
	Jonas Klingwort$^{1,\ast}$ and Vera Toepoel$^{1}$\\
	\\
	\normalsize{$^{1}$Department of Research \& Development, Statistics Netherlands (CBS),}\\ \normalsize{CBS-weg 11, PO Box 4481, 6401 CZ Heerlen, the Netherlands}\\
	\normalsize{$^\ast$correspondence: j.klingwort@cbs.nl}
}
\date{~}

\begin{document}
	
	\maketitle
	
	% ===================================================================
	\begin{abstract}
		\noindent \textbf{Abstract:}
		When conducting a survey, many choices regarding survey design features have to be made. These choices affect the response rate of a survey. This paper analyzes the individual effects of these survey design features on the response rate. For this purpose, data from a systematic review of crime surveys conducted in Germany between 2001--2021 were used. First, a meta-analysis of proportions is used to estimate the summary response rate. Second, a meta-regression was fitted, modeling the relationship between the observed response rates and survey-design features, such as the study year, target population, coverage area, data collection mode, and institute. The developed model informs about the influence of certain survey design features and can predict the expected response rate when (re-) designing a survey. This study highlights that a thoughtful survey design and professional survey administration can result in high response rates.
	\end{abstract}
	
	\small{non-response; victimization; data quality}
	
	\section{Introduction}\label{sec:intro}
	
	\noindent The achieved response rate of a survey depends on several design features, for example, the survey's length and complexity, offered incentives, or the reputation and credibility of the conducting survey agency \parencite{Don2020}. In recent years, declining trends in survey response rates have been observed \parencite{Meyer2015}. Nevertheless, these findings might not be generalized as well-designed surveys can still yield high response rates \parencite{Holtom2022Feb}. Consequently, it is important to know in which direction specific survey design features influence the response rate. With this knowledge, tailored and specific survey design features can be chosen when conducting a population survey.
	
	This work aims to quantify the effects of survey design features on the response rate. This approach is demonstrated using German crime surveys. No regular or repeated standardized official crime survey is conducted in Germany. Instead, criminological survey research in Germany is dominated by individual and independent surveys. These differ, for example, by contractor, executive survey agency, target population, sampling design, or data collection mode. This circumstance allows to study the effects of different survey design features and their impact on the response rate.
	
	For this study, the initial systematic review by \textcite{Klingwort2017} is used but has been updated. The updated systematic review considers German crime surveys between 2001--2021 and collected survey design feature information based on survey-related publications. The survey design features, such as study year, target population, coverage, data collection mode, responsible institute/con\-trac\-tor, sample, and response size, were extracted from these publications. 
	
	First, a meta-analysis of proportions is used to estimate the summary effect size, i.e., the response rate. Second, using a meta-regression, the effects of the survey design features on the response rate are estimated. Moreover, it will be informed on the models' prediction quality, optimal and non-optimal survey design feature sets, and potential model selection.
	
	\section{Background}\label{sec:backg}
	
	\noindent Response rates of population surveys show a decline during the last decades \parencite{Meyer2015}, \parencite{CzajkaBeyler2016}, \parencite{Williams2018Jun}, \parencite{Luiten2020Sep}, \parencite{Daikeler2020Jun}, \parencite{Dutwin2021Jun}, \parencite{Lugtig2022}. For crime surveys even lower response rates can be expected since sensitive topics are surveyed \parencite{Tourangeau1996Jan}. Nevertheless, the number of conducted surveys is enormous and increasing \parencite{Singer2016}. For example, \textcite{PresserMcCulloch2011} reported that the number of surveys conducted between 1984--2004 increased at a rate many times greater than the population in the US. However, in the context of decreasing survey response and increasing costs for data collection, the future of surveys has been questioned \parencite{Couper2013}, \parencite{Alwin2013}, \parencite{Couper2017,} \parencite{RaoFuller2017}, and as a result, attempts and initiatives emerged in recent years to find alternatives to surveys \parencite{Link2014Jan}, \parencite{Galesic2021Jul}. From these developments, we derive the importance of this study. There is an urgent need to ensure that response rates do not decline further and, ideally, start to increase again. In an ideal situation, a survey would be designed so that no missing data would occur \parencite{Allison2002}. However, this ambitious goal can, at best, be approached. A well-designed survey and professional administration are essential in approaching this solution. In this work, we will provide answers and recommendations regarding this solution. 
	
	\textcite{Don2020} identified seven factors related to survey design and administration that influence survey response rates. First, the choice of data collection mode or combination of data collection modes affects the response rate. For example, CATI surveys achieve only low response rates \parencite{Kohut2012}, \parencite{Habermann2017}, \parencite{Olson2020}. In recent years, web surveys have become more prominent in survey research. However, web surveys also commonly achieve lower response rates than other modes \parencite{Blumenberg2018}, \parencite{Daikeler2019}. Due to the particular advantages and disadvantages of different modes, their combinations, and advancing technical development, there have been developments in adaptive and responsive survey designs \parencite{Groves2006}, \parencite{Axinn2011}, \parencite{Schouten2013}, \parencite{Luiten2013}, \parencite{Laflamme2016}, \parencite{Chun2018Sep}, mixed-mode or multiple-mode surveys \parencite{de2005}, \parencite{Greenlaw2009Jul}, \parencite{Dillman2009Mar}, \parencite{MillarDillmann2011}, \parencite{DillmannSmythChristian2014}. In an adaptive survey design, e.g., the data collection mode or contact times are adjusted during the fieldwork. Surveys using a responsive design have sequential phases, wherein in each phase, different fieldwork strategies are applied \parencite{Chun2018Sep}. Mixed-mode surveys use different data collection modes for different subgroups, and the respondents in multiple-mode surveys are interviewed at one time with different modes \parencite{Schnell2019}.
	
	Second, the sponsorship of the survey affects the response rate. For example, there is evidence that surveys of governmental bodies and agencies achieve higher response rates than commercial survey agencies \parencite{Brick2012Nov}. Therefore, a negative connotation about the sponsor will result in substantive adverse effects on the response rate \parencite{Faria1996Feb}, \parencite{Edwards2014Jan}, \parencite{Isani2022Sep}. 
	
	Third, the (perceived) response burden of the survey affects the response rate. Here, factors such as the questionnaire length \parencite{Galesic2009Jan}, motivation or perception of the survey \parencite{Yan2020Jan}, and the respondent's satisficing strategies \parencite{Krosnick1996Jun} will affect the response rate. A recent review and conceptual framework on factors contributing to the (perceived) response burden is given by \textcite{Yan2022Dec}.
	
	Fourth, the offered (appropriate) incentives affect the response rate. There is a large body of literature showing that incentives increase response. However, monetary incentives are more effective than gifts. In addition, incentives received in advance work better than those received after survey participation. However, the optimal size of incentives is debated, so the use of tailored incentives is recommended \parencite{Toepoel2012Jun}, \parencite{SingerYe2013}, \parencite{Singer2018}.
	
	Fifth, factors of the fieldwork strategy, such as a larger number of reminders, the number of contact attempts made, or applied refusal conversion affect the response rate positively \parencite{Sturgis2017May}, \parencite{Beullens2018Apr}, \parencite{Klingwort2018}, \parencite{Mcgonagle2022Feb}. 
	
	Sixth, emphasizing the added value of participation through an appealing display of survey content. This includes adequate explanations regarding confidentiality protection, contact options in case of inquiries, and appropriate communication channels \parencite{Don2020}.
	
	Seventh, consider the features and attributes of the target population. Individual attributes, for example, education or gender, affect whether a response is given. Accordingly, these characteristics must be considered when surveying a population, especially if they correlate with the target variables of the survey \parencite{Don2020}.
	
	Of the seven features listed, the data collection mode, the sponsorship and responsible survey conductor, and features of the target population will be considered in the present analysis. In addition, we will use information about the study year and the coverage area. The discussion will address why other aspects of \textcite{Don2020} were omitted.
	
	\subsection{Crime surveys}\label{sec:crimesvys}
	
	\noindent Self‐report survey methods inform about the `dark figures' or `hidden figures' of crime and are therefore important in criminological and social research. Such surveys complement official crime statistics based on official crime cases reported to the police or cases that lead to an arrest. Combining both data sources allows quantifying crime rates and fear of crime \parencite{MaxfieldWeilerWidom2000}. Methodological aspects in such surveys are discussed by \parencite{Kreuter2002}, \parencite{Noack2015}, \parencite{SchnellNoack2015}. Countries such as the USA (National Crime Victimization Survey), the UK (Crime Survey for England \& Wales), and the Netherlands (Safety Monitor) conduct periodically repeated victimization surveys of the general population. In Germany, no regular or repeated standardized official crime survey is conducted. From a criminological and national security perspective, this is an unfortunate state of affairs (addressed in the German national security reports \parencite{Bundesministerium2001}, \parencite{Bundesministerium2006}, \parencite{Bundesministerium2021}). Individual and independent surveys instead dominate crime survey research. This fact enables us to systematically study the effects of survey design features on the response rate. However, the first studies that use big data for crime statistics are published, but these need to be carefully evaluated \parencite{Lohr2019a}. Accordingly, sample surveys seem currently without alternatives.
	
	\section{Study selection}\label{sec:studsel}
	
	\noindent The initial systematic review was done by \textcite{Klingwort2017} and considered the period from 2001--2016. Since then, additional and relevant crime surveys have been conducted. Therefore, the systematic review has been updated for this paper. As the focus is on German victimization surveys, literature research was conducted primarily in German databases and at German institutions. The databases used are: `Deutsche Nationalbibliothek', `Gemeinsamer Verbundkatalog', `Sociological Abstracts', `Sowiport', `Web of Science', `WISO', `Google Scholar', and `KrimDok'. Furthermore, the websites and archives of the following institutions and ministries were queried: `Bundeskriminalamt', `Landeskriminalämter', `Bundesministerium für Familie, Senioren, Frauen und Jugend', `Bundesministerium für Gesundheit', `Bundesministerium des Innern', `Statistisches Bundesamt', `Statistische Landesämter', `Robert Koch-Institut', `Leibniz-Institut für Sozialwissenschaften', `Deutsches Institut für Wirtschaftsforschung', `Kriminologisches Forschungsinstitut Niedersachsen', `Eurostat', `European Union Agen\-cy for Fundamental Rights', `Deutsche Forschungsgemeinschaft'.
	
	The following German search terms were used: `Opferbefragung (*)', `Vik\-ti\-mi\-sie\-rungs\-sur\-vey (*)', `Viktimisierungsbefragung (*)', `Kriminalitätsfurcht', `Befragung zu Kriminalitätsfurcht', `Befragung zu Wohnungseinbruch', `Befragung zu Sicherheit', `Befragung zu Kriminalität', `Befragung zu Gewalt'. Furthermore the references in the researched sources were searched for further eligible studies. The criteria that the surveys must meet to be included are: 
	
	\begin{enumerate}
		\item Population: Germany
		\item Field period: 01.01.2001--31.12.2021
		\item Gross sample size: $n \gtrapprox$ 1000
		\item Questions about forms of physical violence and/or, victimization in the form of residential burglary and/or, questions on fear of crime
		\item (Gross) sample size, (net) response size, and response rate
	\end{enumerate}
	
	\noindent Population 'Germany' means that the sample is based on a sampling frame with addresses/contact information within Germany. There are no restrictions regarding German citizenship or the German-speaking population. The chosen field period connects with some overlap to the last overview published by \textcite{Obergfell-Fuchs2008} that listed the victimization surveys conducted in Germany between 1973 and 2005. The gross sample size was set $n \gtrapprox$ 1000 to prevent a potentially large number of surveys with small sample sizes from being included and to consider only the larger and more relevant surveys. How many more surveys would have been included without this restriction is unknown. However, we explain in this section that two surveys with a smaller gross sample size were included.
	
\begin{figure}[htbp]
		\includegraphics[width=\linewidth]{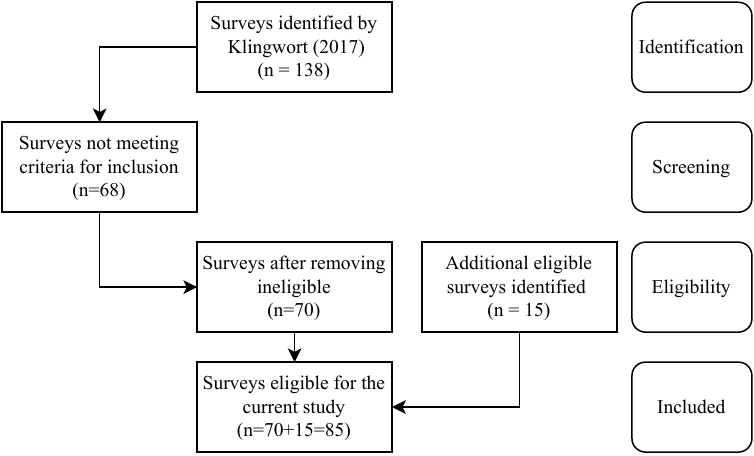}
		\caption{PRISMA search strategy to identify eligible surveys}
		\label{fig:XXXX_fig1}
	\end{figure}

	Figure \ref{fig:XXXX_fig1} shows how the eligible surveys were identified (according to PRISMA guidelines, see \url{https://prisma-statement.org/}). \textcite{Klingwort2017} identified $138$ different samples. Note that the number of samples does not correspond to the number of studies. It is possible that more than one independent sample has been drawn within one study, which, for example, considered different target populations. In this case, all independent samples of that study have been considered eligible and the response rates were calculated for each individual sample.
	
	Of the $138$ different samples, $68$ samples are considered not eligible for the current study. Those are either quota samples not allowing to calculate a response rate, other forms of non-probability samples, a long-term panel, or other general population surveys without a focus on crime. In two cases, drop-off questionnaires were used, which are not considered in this study because they are not based on an independent sample. Two included studies are below the targeted gross sample size but included because they belong to a study in which an additional independent larger sample was drawn that qualified to be included. 
	
	For the current study, additional $15$ samples were identified. These studies were identified using the same strategy as for the initial systematic review by \parencite{Klingwort2017}. However, only the number of eligible surveys is documented. As a result, $n=85$ samples are eligible and included. 
	
	\noindent The publications for the identified studies were screened, and reported information about the (gross) sample size, (net) response size, response rate, and the survey design features (study year, target population, coverage area, data collection mode\footnote{CATI: Computer Assisted Telephone Interview, CAWI: Computer Assisted Web Interview, F2F: Face-to-face interview, PAPI: Paper assisted personal interview}, and institute) were documented. In some cases, only the response size and the response rate were reported. Here, the gross sample size was calculated as $\round{\frac{\rm \rm{response ~size ~* ~100}}{\rm{response ~rate ~(in ~\%)}}}$. The documented variables are described in the following. 
	
	\subsection{Response rates}\label{sec:rr}
	
	\noindent We have calculated the response rates using the sample and response size to ensure that the response rates are the same type. By this, we avoid having differently calculated response rates in the target variable. The distribution of the response rates of the samples used is reported in Table \ref{tab:rr}. The set `All' is based upon all 85 samples with an average response rate of 0.580 and 0.127 as minimum and 0.941 as maximum. Of those 85 samples, 40 are based upon `classroom interviews/School-panels'. Given that this design differs considerably from a cross-sectional design (Csd), we also report the response rate distribution solely for the 45 Csd samples. For the Csd studies, an average response rate of 0.412 is observed, with 0.127 as the minimum and 0.860 as the maximum. In both sets, the mean and median are close, indicating symmetrical distributions (see also Section \ref{sec:metaprop}). Both sets will be considered for the meta-analysis of proportions, see Section \ref{sec:res}.
	
	\begin{table}[h]
		\caption{Distribution of response rates split by different sets of data}
		%\resizebox{\linewidth}{!}{
			\centering
			\begin{tabular}{lcccccc}
				\toprule
				Set &  Min.   & $Q_{25}$ &  $Q_{50}$ &    Mean & $Q_{75}$ &  Max.\\
				\midrule
				All (n=85) &  0.127  & 0.390 &  0.569  & 0.580 & 0.828 & 0.941  \\
				Csd (n=45) & 0.127  & 0.270 &  0.395  & 0.412 & 0.511 & 0.860  \\
				\bottomrule
			\end{tabular}
			%   }
		\label{tab:rr}
	\end{table}
	
	\subsection{Survey design features and coding}\label{sec:svyfeat}
	
	\noindent Table \ref{tab:features} shows the documented survey design features, their distribution and categorization.
	
	\begin{table}[h]
		\caption{Distribution and categorization of survey design features}
		%\resizebox{\linewidth}{!}{% <------ Don't forget this %
			\centering
			\begin{tabular}{ll}
				\toprule
				Feature & Distribution and categorization \\
				\midrule
				Study year & Min: 2001, Mean: 2010, Max.: 2020\\
				Target population & General population (36); \\
				& Non-general population (49) \\
				Coverage area & National level (20); Regional level (26);\\ 
				& Local level (39) \\ 
				Data collection mode & CATI (14); CAWI (4); F2F (6); PAPI (21); \\
				& Classroom interviews/School-panels (40)\\
				Institute & Public institute (10); Ministry (12); \\
				& National criminological institute (32); \\
				& Police (15); University (16)\\
				\bottomrule
			\end{tabular}
			%   }
		\label{tab:features}
	\end{table}
	
	\noindent To avoid sparse categories, efforts were made to code in such a way that at least five surveys could be assigned to a category. For CAWI, only four studies were found. These were not pooled with other categories. The starting year was selected if the study did not finish within a year. Although the period till 31.12.2021 is considered, the maximum for the study year is 2020. Only two surveys used mixed-mode, that were assigned to PAPI, which was used as the main/first mode. Classroom interviews and school-panels were pooled. If several institutes were involved, the first listed was considered. International and national public institutes were pooled. For coverage area, the national level refers to studies conducted country-wide, the regional level refers to studies conducted on the federal-state level, and the local level refers to studies conducted in cities or municipalities.
	
	\section{Methods}
	
	\subsection{Meta-analysis of proportions}\label{sec:metaprop}
	
	\noindent A meta-analysis of proportions considers a proportion as effect size, such as the proportion of people in a study who experienced a particular outcome, e.g., recovering from a disease \parencite{Borenstein2009}, \parencite{Schwarzer2022}. Transferred to the current study, we aim at synthesizing response rates from multiple studies to provide a precise estimate of the true effect size, i.e., the response rate. For the current study, it is reasonable to assume that the studies do not stem from a single population, and therefore, a random-effects model will be fitted that assumes a distribution of true effect sizes. Hence, the mean of the distribution of true effect sizes is estimated \parencite{Borenstein2009}, \parencite{Harrer2022}. The random-effects model can be expressed as
	
	\begin{equation}
		\hat{\theta}_k = \mu + \zeta_k + \epsilon_k,
	\end{equation}
	
	\noindent with $\hat{\theta}_k$ being the observed effect size of study $k$, $\mu$ the true effect sizes mean, and the two error sources $\zeta_k$ (true variation in
	effect sizes) and $\epsilon_k$ (sampling error) \parencite{Borenstein2009}, \parencite{Harrer2022}. A weighted mean is computed to obtain a precise estimate of $\mu$, with the weights based on the inverse of a study’s variance (sum of the within-study variance and the between-studies variance). The within-study and between-studies variance are necessary to compute a study's variance. Therefore, different estimators exist, with differences in estimating the between-study variance. In the current study, the restricted maximum-likelihood estimator (REML) is used to estimate the between-study variance \parencite{Thorlund2011Dec}. The REML is a suitable estimator for continuous outcome data. For details, we refer to \textcites{Borenstein2009,Chen2013,Harrer2022,Evangelos2022}. Other available estimators were also tested, such as DerSimonian-Laird, Paule-Mandel, Empirical Bayes, and Sidik-Jonkman. None of these estimators would have yielded substantially different results than those presented in Table \ref{tab:modelres}. The results of this analysis are not shown.
	
	It might be required to apply transformations to the considered proportions to improve their statistical properties regarding following a normal distribution. Usually, the logit or double-arcsine transformation is used \parencite{Schwarzer2022}. However, when proportions around 0.5 are observed, and the number of studies is large, it can be assumed the proportions follow a binomial distribution. In such a case, no transformations are required because the normal distribution approximates the binomial distribution. Moreover, when the observed proportions are between 0.2 and 0.8, no transformations are required \parencite{Lipsey}. Considering the response rate distributions reported in Table \ref{tab:rr}, the observed proportions will be used, and no transformations will be applied.
	
	To quantify heterogeneity, we report several metrics. First, Cochran’s $Q$ is used to disentangle the sampling error and the between-study heterogeneity. It is used to test if the heterogeneity exceeds the one expected under the null (no heterogeneity). Second, the $I^2$ statistic, which is the percentage of variability, that is not caused by sampling error. Finally, the heterogeneity variance ($\tau^2$) and the heterogeneity standard deviation ($\tau$). The variance in the true effect sizes is quantified by $\tau^2$. From that, $\tau$ is obtained and is expressed in the same metric as the effect sizes \parencite{Borenstein2009}, \parencite{Harrer2022}. 
	
	\subsection{Meta-regression}\label{sec:metareg}
	
	\noindent A meta-regression assumes a mixed-effects model and accounts for the study's deviation from the true effect because of between-study variance and sampling error. In a meta-regression study, design features that may have influenced the results can be used. Hence, a meta-regression predicts the observed effect size $\hat{\theta}_k$ \parencite{Harrer2022}. The model is expressed as
	
	\begin{equation}
		\hat{\theta}_k = \theta + \beta x_k + \epsilon_k + \zeta_k,
	\end{equation}
	
	\noindent with $\beta$ the regression coefficient and $x$ the predictor (study design feature) of study $k$. For details, we refer to \textcites{Borenstein2009,Harrer2022}. Accordingly, the model in the present study will be based on $k=85$ data points. Given the limited number of data points, the regression will be restricted to main effects only. For the meta-regression, $\tau^2$, $\tau$, $I^2$, will be reported as well. Moreover, $R^2_*$ is reported, which considers the residual heterogeneity variance not explained by the meta-regression slope and relates it to the total heterogeneity. Finally, residual heterogeneity and moderators are tested.
	
	\noindent All analyses were conducted using \texttt{R, version 4.1.3} \parencite{Rcore2022}. For the meta-analyses, the \texttt{R}-library \texttt{metafor} was used \parencite{Viechtbauer2010}.
	
	\section{Results}\label{sec:res}
	
	\noindent The estimates of the summary proportion are shown in Table \ref{tab:modelres}. For the set `All,' a summary proportion of 0.580 with an SE of 0.026 and a CI between 0.529 and 0.631 is estimated. For the set `Csd,' a summary proportion of 0.412 with an SE of 0.025 and a CI between 0.364 and 0.461 is estimated. Thus, the random-effects model with the REML estimator yields similar estimates for the summary proportion of the response rate as the means shown in Table \ref{tab:rr}.
	
	\begin{table}[htbp]
		\centering
		\caption{Summary proportion of the response rate (true effect size), standard error (SE), and 95\% confidence interval (CI) split by different sets of data.}
		\begin{tabular}{lrrr}
			\toprule
			\multicolumn{1}{l}{Set} & \multicolumn{1}{c}{Estimate} & \multicolumn{1}{c}{SE} & \multicolumn{1}{c}{CI} \\
			\midrule
			All & 0.580  & 0.026 & [0.529; 0.631]  \\ 
			Csd & 0.412 & 0.025 & [0.364; 0.461]  \\ 
			\bottomrule
		\end{tabular}
		\label{tab:modelres}
	\end{table}
	
	\noindent The heterogeneity test for Set `All' shows that $Q = 346676.8753$, which is larger than expected with df = 84. Accordingly, the test for heterogeneity is significant ($p < .05$). For the set `Csd', $Q = 74352.5739$ and larger than expected with df = 44. The test for heterogeneity in this set is also significant ($p < .05$). Table \ref{tab:reshetero} shows the additional metrics to quantify heterogeneity. First, the results for set `All' are considered. The between-study heterogeneity variance ($\tau^2$) is estimated at 0.058 (95\% CI: 0.044 - 0.081). The true response rates (effect sizes) have an estimated standard deviation at $\tau = 0.242$ (95\% CI: 0.210 - 0.285). The $I^2$ statistic is 99.979\% (95\% CI: 99.972 - 99.985), indicating substantial heterogeneity. Second, $\tau^2$ for the set `Csd' is estimated at 0.028 (95\% CI: 0.019 - 0.044) and $\tau = 0.167$ (95\% CI: 0.138 - 0.211). The $I^2$ statistic is 99.967\% (95\% CI: 99.952 - 99.980), indicating substantial heterogeneity. Moreover, the confidence intervals for $\tau^2$ of both sets do not contain zero indicating between-study heterogeneity. Summarizing the results from these metrics, there is strong evidence that true response size differences cause variation in the data. The extent to which the considered survey design features are able to explaining the variation in the data will be reported in the next section.
	
	In an additional analysis, we checked whether these reported results and conclusions would change if logit-transformed proportions had been used. No substantial differences were found. The estimates in Table \ref{tab:modelres} would vary by 1-2 percentage points. The heterogeneity measures reported in Table \ref{tab:reshetero} would yield the same conclusions.
	
	\begin{table}[htbp]
		%\resizebox{.5\textwidth}{!}{% <------ Don't forget this %
			\caption{Metrics to quantify heterogeneity split by different sets of data}
			
			\centering
			\begin{tabular}{lrrr}
				\toprule
				All &   Estimate  &   \multicolumn{1}{c}{CI} \\
				\midrule
				$\tau^2$   &   0.058 &   [0.044;  0.081] \\
				$\tau$      &  0.242  &  [0.210; 0.285] \\
				$I^2 ~ (\%)$  &  99.979 &  [99.972; 99.985] \\
				\midrule
				Csd \\
				\midrule
				$\tau^2$   &  0.028   &  [0.019; 0.044] \\
				$\tau$     &  0.167   &  [0.138; 0.211] \\
				$I^2 ~ (\%)$ &  99.967  &  [99.952; 99.980] \\
				\bottomrule
			\end{tabular}
			%   }
		\label{tab:reshetero}
	\end{table}
	
	\subsection{Meta-regression}\label{sec:res3}
	
	\noindent Table \ref{tab:resmeta} shows the results of the meta-regression. A permutation test for the p-values is used to validate the robustness of the model. By this, it is possible to assess better whether the coefficients capture a true pattern or if the model picks up statistical noise (see also \textcite{Harrer2022}). 
	
	\begin{table}[h!]
		\caption{Results of the Meta-regression model. The estimated coefficients, standard error (SE), and confidence interval (CI) are shown. The $p$-values are based on a permutation test with 1000 iterations.}
		\resizebox{\linewidth}{!}{% <------ Don't forget this %
			\centering
			\begin{tabular}{lrrrr}
				\toprule
				& \multicolumn{1}{c}{$\beta$} & \multicolumn{1}{c}{SE} & \multicolumn{1}{c}{$p$} & \multicolumn{1}{c}{CI} \\
				\midrule
				Intercept                                 &  22.835 & 6.967 &  0.002 &  [9.179; 36.491] \\
				\\
				Year                                    &  -0.011 & 0.003 &  0.002 & [-0.018; -0.004] \\
				\\
				Population (reference: general population) \\
				~~ Non-general pop.                        &  -0.045 & 0.052 &  0.404 & [-0.146; 0.056] \\  
				\\
				Coverage area (reference: local) \\
				~~ National                        &   0.103 & 0.051 &  0.055 &  [0.004; 0.203] \\  
				~~ Regional                        &   0.022 & 0.040 &  0.608 & [-0.057; 0.101] \\   
				\\
				Data collection (reference: CATI) \\
				~~ CAWI                                &   0.164 & 0.083 &  0.044 &  [0.001; 0.327] \\
				~~ Classroom interviews/  &   0.444 & 0.076 &  0.001 &  [0.296; 0.593] \\
				~~ School-panels \\
				~~ F2F                                 &   0.137 & 0.081 &  0.088 & [-0.022; 0.296] \\ 
				~~ PAPI                                &   0.201 & 0.051 &  0.001 &  [0.101; 0.301] \\
				\\
				Institute (reference: Ministry) \\
				~~ National crim. institute  &   0.049 & 0.057 &  0.413 & [-0.063; 0.161] \\ 
				~~ Police                             &  -0.038 & 0.055 &  0.495 & [-0.145; 0.069] \\   
				~~ Public institute                   &  -0.075 & 0.063 &  0.222 & [-0.199; 0.048] \\   
				~~ University                         &   0.180 & 0.069 &  0.013 &  [0.045; 0.315] \\ 
				\midrule
				$k = 85$ \\
				$\tau^2 = 0.017 ~ (\text{SE} = 0.003)$ \\ 
				$I^2 = 99.900$ \\
				$R^2_* = 70.310$ \\
				$QE = 67271.114, p < 0.05$ \\
				$QM = 210.453, p < 0.05$ \\
				\bottomrule
			\end{tabular}
		}
		\label{tab:resmeta}
	\end{table}
	
	\noindent The intercept is estimated at 22.835 (95\% CI: 9.179 - 36.491) with an SE of 6.967. The feature `Year' has a negative coefficient (-0.011), and the CI indicates only negative effects of this feature. Accordingly, on average, each progressing year causes a decrease in the response rate of -0.011. The p-value of 0.002 indicates statistical significance, i.e., the null hypothesis can be rejected, and the coefficient is significantly different from 0. Considering the target population, studies surveying non-general populations score on average a -0.045 lower response rate compared with those surveying the general population. The p-value of 0.404 indicates no statistical significance, i.e., the null hypothesis cannot be rejected, and the coefficient is not significantly different from 0. When the survey is conducted nationwide or on a regional level, in both cases, average larger response rates are achieved than with locally conducted surveys. The p-values of 0.055 and 0.608 indicate no statistical significance. All data collection modes score, on average, higher than CATI. For example, using PAPI instead of CATI will, on average, result in an 0.201 higher response rate. When conducting a classroom interview/school-panel, a 0.444 larger response rate can be achieved on average. The p-values of all coefficients indicate statistical significance, except the coefficient for `F2F' (0.088). The coefficients for the feature `Institute' shows that in surveys in which the Police and Public Institutes occupy a prominent function, on average lower response rates are achieved in contrast to surveys in which a Ministry holds a central role. National criminological institutes and Universities achieve, on average, a higher score than Ministries. Here, only the p-value for the coefficient for `University' (0.013) indicates statistical significance.
	
	The variance that the predictors do not explain in this model is $\tau^2 = 0.017 ~ (SE = 0.003)$. According to $I^2$, after the inclusion of the predictors, 99.900\% of the variability is due to the remaining between-study heterogeneity. The $R^2_*$ indicates that the predictors explain 70.310\% of the difference in response rates. The test for residual heterogeneity is significant ($p < 0.05$), with $QE = 67271.114$, meaning that the heterogeneity not explained by the predictors is significant. The test of moderators is significant ($p < 0.05$), with $QM = 210.453$, meaning that some of the included predictors influence the study's response rate.
	
	\subsection{Model predictions}
	
\begin{figure}[h!]
		\centering
		\includegraphics[width=.9\linewidth]{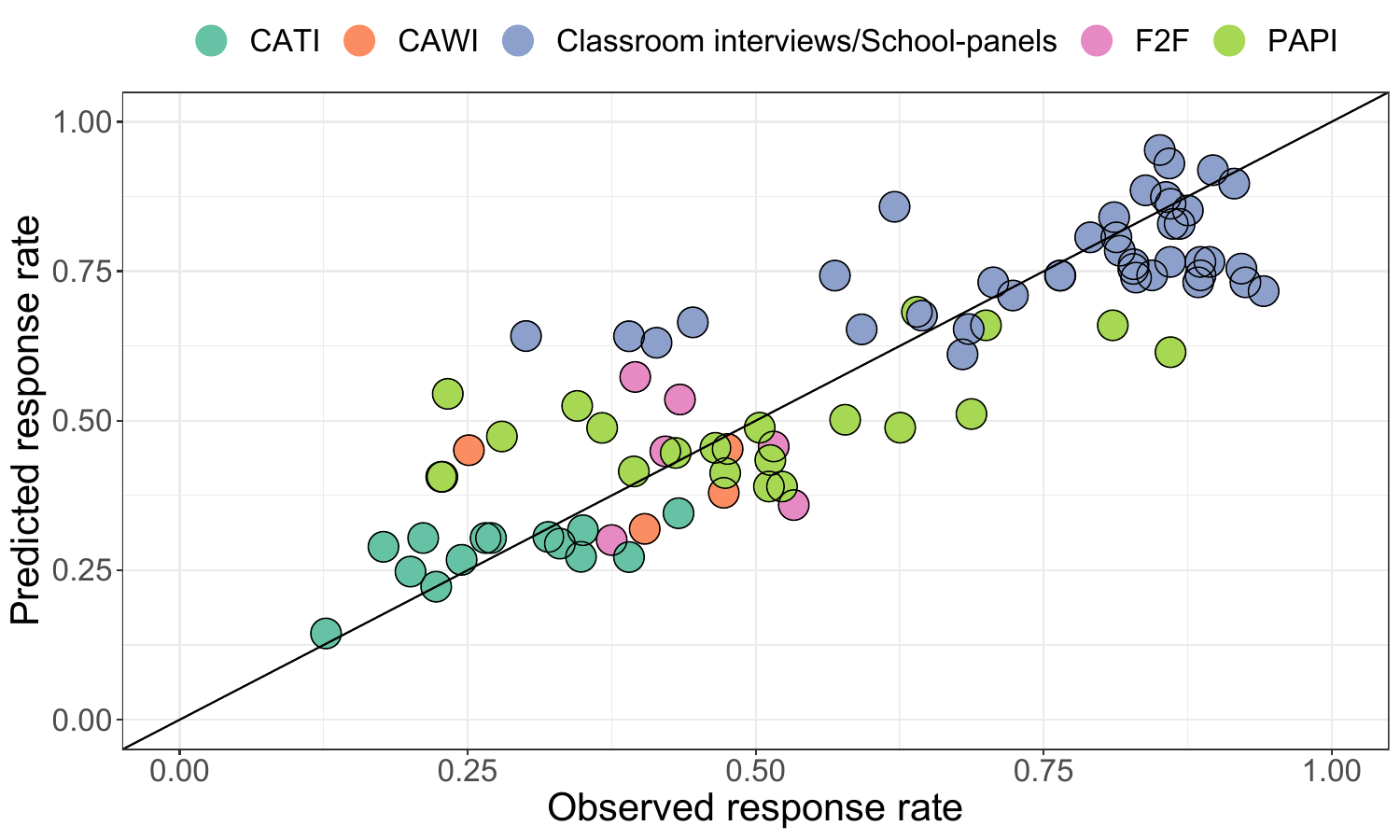}
		\caption{Observed and predicted response rates. The color indicates the data collection mode. The black diagonal shows the regression line.}
		\label{fig:obs_vs_pred}
	\end{figure}

	\noindent Figure \ref{fig:obs_vs_pred} shows the observed and predicted response rate using the observed data and the developed model. The data points are colored by data collection mode and scattered around the regression line, indicating a strong positive correlation ($r=0.863$). 
	
	\noindent Table \ref{tab:obs_vs_pred_diff} shows the distribution of the absolute prediction errors (observed - predicted). On average, the error is 0. The median is close to zero, indicating a symmetrical distribution. The minimum and maximum indicate large prediction errors of -0.341 (overestimation) and 0.245 (underestimation). The MSE = 0.015 and the RMSE = 0.121.
	
	\begin{table}[h!]
		\caption{Distribution of differences between observed and predicted response rates.}
		
		\centering
		\begin{tabular}{lcccccc}
			\toprule
			Min.   & $Q_{25}$ &  $Q_{50}$ &    Mean & $Q_{75}$ &  Max.\\
			\midrule
			-0.341 & -0.047 &  0.015 &  0.000 &  0.078 & 0.245   \\ 
			\bottomrule
		\end{tabular}
		\label{tab:obs_vs_pred_diff}
	\end{table}
	
	The developed model can be used for decision-making when (re-) designing a survey. In Table \ref{tab:designs}, we report the five survey designs with the highest response rates and the five with which the lowest response rates are predicted. Here we do not consider `Classroom interviews/School-panels' as data collection mode and `University' as institute. This decision is due to the use of main effects only. A detailed explanation is given in the limitations of this study, see Section \ref{sec:discuss}). The year of conducting the survey is constant in the different designs and is set to 2024. Increasing the year would result in lower response rates, given the negative coefficient.
	
	The results provide evidence that higher response rates are achieved in nationwide surveys where the general population is considered. The use of PAPI for data collection appears to be superior, and CATI surveys will achieve low response rates. With a criminological institute as the responsible institute, the highest response rates can be expected, while surveys conducted by the police or public institutes will achieve low response rates when using CATI.
	
	\begin{table}[htbp]
		\caption{Selection of survey design feature sets and their predicted response rate (RR), standard error (SE), and confidence interval (CI).}
		
		\resizebox{\linewidth}{!}{
			\centering
			\begin{tabular}{llllccc}
				\toprule
				Population & Coverage & Data collection & Institute & RR & SE & CI\\
				\midrule 
				General & National & PAPI & Crim. institute & 0.457 & 0.068 & [0.325, 0.590] \\
				General & National & CAWI & Crim. institute & 0.420 & 0.097 & [0.230, 0.610] \\
				Non-general & National & PAPI & Crim. institute & 0.413 & 0.089 & [0.240, 0.586] \\
				General & National & PAPI & Ministry & 0.409 & 0.081 & [0.250, 0.567] \\
				General & National & F2F & Crim. institute & 0.393 & 0.095 & [0.207, 0.580] \\
				\midrule
				Non-general & Regional & CATI & Police & 0.043 & 0.079 & [-0.112, 0.197] \\
				General & Local & CATI & Public institute & 0.028 & 0.091 & [-0.149, 0.206]\\
				Non-general & Local & CATI & Police & 0.021 & 0.086 & [-0.148, 0.189]\\
				Non-general & Regional & CATI & Public institute & 0.006 & 0.103 & [-0.196, 0.207] \\
				Non-general & Local & CATI & Public institute & -0.016 & 0.106 & [-0.223, 0.190]\\
				\bottomrule
			\end{tabular}
		}
		\label{tab:designs}
	\end{table}
	
	\noindent For the predicted response rates reported, standard errors between 0.068 -- 0.106 are found. The range of the reported confidence intervals can be considered large. This is likely due to the small number of data points in the model. Moreover, some negative values are predicted. This is due to modeling the original proportions. In these scenarios, the response rate falls between 0 and the upper limit of the confidence interval.
	
	\subsection{Model selection}
	
	\noindent Meta-regressions are usually based on little observations limiting the complexity of the statistical model. Choosing a too-complex model will result in overfitting and imprecise estimates. We aimed to find a probably more sparse model than the model reported in Section \ref{sec:res3} that might still explain a comparable amount of heterogeneity. Therefore, we fitted all potential model combinations (considering main effects only) and compared the obtained $R^2_*$. When leaving out the population feature, an $R^2_* = 70.41\%$ was obtained. Hence, this might be considered a feature to be dropped when the number of included studies is limited. The worst model ($R^2_* = 2.9\%$) was obtained using only the study year. The feature itself, strictly speaking, is not an actual survey design feature.
	
	\section{Discussion}\label{sec:discuss}
	
	\noindent This section highlights the study findings, addresses its limitations, and gives recommendations for future research.
	
	First, the results of the meta-analysis of proportions showed an estimated summary response rate of 58\% and 41.2\% when considering cross-sectional designs only. Second, the results of the meta-regression showed that time has a negative effect on the response rate. Surveying the non-general population has a negative effect as well. Surveys on the national or regional level score higher response rates than locally conducted surveys. All data collection modes score higher than CATI. The findings for the effect of the responsible institute are mixed. Some institutes score higher than ministries (national criminological institute, university), and others lower (police, public institutes). The model predictions based on the observed data are of reasonably good quality. However, some of the predictors were not significant. When the model is applied to new data, results suggest using nationwide general population PAPI surveys. It is not recommended to survey non-general populations with CATI on a local or regional level, with the police or a public institute being responsible. 
	
	The developed model is easy to adopt, given the limited number of features and their resolution. However, the sparsity of the model also implies some limitations. 
	
	First, given the number of data points, the precision of the obtained point estimates is limited. This effect is evident for the estimate precision in Table \ref{tab:modelres}, where the CI ranges are 10\%. This problem becomes even more evident in Table \ref{tab:resmeta}. The CIs of the coefficients are large and do sometimes not allow clear conclusions in one direction. If the obtained precision is acceptable might be related to the research question or the quality standards. This problem can only be solved by including more studies (data points). A study not limited to crime surveys should be able to find more observations more easily.
	
	Second, although the model explains a significant part of the heterogeneity (about 70\%) in the response rates, about a third is not explained, which still contains a significant part of heterogeneity. This is because several potentially important features are not included in the model. For example, the fieldwork duration and strategy, recruitment letter and the number of reminders, survey length, incentives, presence of an interviewer, or interviewer workload. This information could not always be retrieved from the publications or was reported too vague. However, as mentioned, the number of data points is limited, and even if all these features had been available, not all could likely have been included in the model, and model selection would have been required. 
	
	Third, we considered only the main effects while using interactions would have been desirable. For example, the institute's case has a problem of using only main effects. The category `University' has a large positive coefficient. However, all observations with `University' also fall into the category `Classroom interviews/School-panels', which explains the large coefficient. This result is thus somewhat misleading and would be explained by interactions.
	
	Fourth, the generalization of this model to other data and non-crime surveys. It is questionable to what extent the cultural context (e.g., the use of PAPI in Germany) can be transferred to other surveys. However, replication in other countries or with non-crime surveys would provide information about the model's generalizability.
	
	In contrast to several published studies, recent research suggests increasing responserates \cite{Holtom2022Feb}. Considering 1014 surveys from 2010 to 2020 a steadily increasing average response rate is reported: 48\% (2005), 53\% (2010), 56\% (2015), and 68\% (2020). Considering these findings and the findings of our work, we see these results as an encouragement for survey researchers to work towards well-suited survey designs that will result in higher response rates. 
	
	\section{Conclusion}
	
	\noindent We studied the effect of survey design features (nonsampling errors) on the response rates in crime surveys. Heterogeneity in response rates has been observed that cannot be explained by a random process. The survey design features (study year, target population, coverage area, data collection mode, and responsible institute) explained a large part of the observed heterogeneity. Results highlight the need for an appropriate survey design and professional administration. 
	
	\section*{Reference list of included studies}
	
	\begin{refsection}
		
		\noindent The reference list below does not contain 85 references (number of eligible samples/surveys) because within some studies $>1$ survey were conducted, respectively independent samples drawn or a reference contains results for several surveys (see also Section \ref{sec:studsel}). 
		
		\nocite{LudwigKraupl2005, Landeskriminalamt2006, MullerSchrottle2005, BaierEtal2011, GörgenEtal2009, DijkEtal2005, BrettfeldWetzels2007, Enzmann2010, BaierEtal2006, RaboldEtal2008, BaierEtal2009a, BaierEtal2009b, Landeskriminalamt2015, Baier2011, Schlack2013, BaierEtal2010, Infas2009, European2014a, BringsEtal2010, BaierRabold2012, ICVS2010, Liebl2014, WollingerEtal2014, Infas2004, DoeringBaier2011, EllrichEtal2012, BaierBergmann2013, BaierPfeiffer2011, GoergenEtal2013, SchielEtal2013, European2014b, JagerEtal2013, EllrichBaier2014, LKANiedersachsen2015, KronebergEtal2016, BergmannEtal2017, Baier2015, AllroggenEtal2016, LKANiedersachsen2016, Dreißigacker2016, Starcke2019, Dreißigacker2017, BergmannEtal2019, BirkelEtal2019, KriegEtal2019, KriegEtal2020, Birkel2020, MotzkeBrondies2004, Brondies2004, Hilfert2005, Kunadt2006, Bentrup2007, Bentrup2009, Bentrup2010, Bentrup2012, BentrupVerneuer2014, Verneuer2015, Verneuer2017, Kessler2019, Kessler2021}
		
		\printbibliography[heading=subbibliography]
		
	\end{refsection}
	
	%\newpage
	\section*{Bibliography}
	
	\printbibliography
	
	\section*{Acknowledgements}
	
	%\noindent We thank two anonymous reviewers for their time, effort, and comments that improved the manuscript.
	
	\noindent The views expressed in this report are those of the authors and do not necessarily correspond to the policies of Statistics Netherlands.
	
\end{document}